\documentclass[usenatbib,usegraphicx]{mn2e}
\usepackage{amsmath}
\usepackage{amssymb}
\usepackage{epsfig}
\usepackage{verbatim}

\newcommand\lsim{~\lower.5ex\hbox{$\buildrel < \over \sim$}~}
\newcommand\gsim{~\lower.5ex\hbox{$\buildrel > \over \sim$}~}

\title{The Radio Signatures of the First Supernovae}

\author[A. Meiksin, D. Whalen]{Avery Meiksin${}^{1}$, Daniel J. Whalen${}^{2}$ \\
SUPA\thanks{Scottish Universities Physics Alliance},
${}^1$Institute for Astronomy, University of Edinburgh,
Blackford Hill, Edinburgh\ EH9\ 3HJ, UK\\
${}^2$McWilliams Fellow, Department of Physics, Carnegie-Mellon University,
Pittsburgh, PA 15213, USA}

\begin{document}

\maketitle

\begin{abstract}
  Primordial stars are key to primeval structure formation as the
  first stellar components of primeval galaxies, the sources of cosmic
  chemical enrichment and likely cosmic reionization, and they
  possibly gave rise to the super-massive black holes residing at the
  centres of galaxies today. While the direct detection of individual
  Pop III stars will likely remain beyond reach for decades to come,
  we show their supernova remnants may soon be detectable in the
  radio. We calculate radio synchrotron signatures between
  $0.5-35$~GHz from hydrodynamical computations of the supernova
  remnants of Pop III stars in $\sim10^7{\rm M_\odot}$ minihaloes. We
  find that hypernovae yield the brightest systems, with observed
  radio fluxes as high as $1-10\,\mu$Jy. Less energetic Type II
  supernovae yield remnants about a factor of 30 dimmer and
  pair-instability supernova remnants are dimmer by a factor of more
  than 10,000. Because of the high gas densities of the progenitor
  environments, synchrotron losses severely limit the maximum emission
  frequencies, producing a distinctive peaked radio spectrum
  distinguishable from normal galactic supernova remnant
  spectra. Hypernovae radio remnants should be detectable by existing
  radio facilities like eVLA and eMERLIN while Type II supernova
  remnants will require the Square Kilometre Array. The number counts
  of hypernova remnants at $z>20$ with fluxes above $1\,\mu$Jy are
  expected to be one per hundred square degree field, increasing to a
  few per square degree if they form down to $z=10$. The detection of
  a $z>20$ Type II supernova remnant brighter than 1~nJy would require
  a 100--200 square degree field, although only a 1--2 square degree
  field for those forming down to $z=10$. Hypernova and Type II
  supernova remnants are easily separated from one another by their
  light curves, which will enable future surveys to use them to
  constrain the initial mass function of Pop III stars.
\end{abstract}


\section{Introduction}

The cosmic Dark Ages ended with the formation of the first stars in
$10^5 - 10^6\, {\rm M_\odot}$ cosmological haloes at $z \sim 20 - 30$.
Primordial (or Pop III) stars are the key to understanding primeval
galaxies, the onset of early cosmological reionization and chemical
enrichment, and the origins of the supermassive black holes found in
most massive galaxies today. Unfortunately, in spite of their extreme
luminosities \citep{s02} and the arrival of next-generation near
infrared (NIR) observatories such as the \textit{James Webb Space
  Telescope} (\textit{JWST}) and the Thirty-Meter Telescope (TMT),
individual Pop III stars will remain beyond the reach of direct
detection for decades to come. For now, there are no observational
constraints on either their masses or their rates of formation.

On the simulation frontier, there has been a gradual shift in paradigm
over the past decade from single $30 - 300\, {\rm M_\odot}$ stars
forming in isolation in haloes \citep{bcl02,nu01,abn02,
  on07,2008ApJ...673...14O,2007ApJ...671.1559W} to binaries
\citep{turk09} and more recently to the possibility of $20 - 40\, {\rm
  M_\odot}$ stars forming in small multiples of up to a dozen
\citep{stacy10,clark11,get11,hos11, stacy11,get12}. However, these
simulations do not estimate Pop III stellar masses by modeling the
formation and evolution of the stars. Instead, they usually derive
them by comparing infall rates at the center of the halo at very early
stages of collapse to Kelvin-Helmholz contraction times to place upper
limits on the final mass of the star. No simulation currently bridges
the gap between the initial formation and fragmentation of a
protostellar disk and its photoevaporation up to a million years
later, so there are no firm numerical constraints on the Pop III IMF
either. \citep[See][ for a recent review of primordial star
  formation.]{whalen12}.

Pop III stars profoundly transformed the haloes that gave birth to
them, driving out most of their baryons in supersonic ionized outflows
\citep{wan04,ket04,abs06,awb07} and later exploding as supernovae
(SNe) \citep{byh03,ky05,get07,2008ApJ...682...49W}. Ionization fronts
from these stars also engulfed nearby haloes, either promoting or
suppressing star formation in them and regulating the rise of the
first stellar populations
\citep[e.g.][]{2004MNRAS.348..753S,il05,su06,
  2008ApJ...679..925W,2009MNRAS.395.1280H,suh09,2010ApJ...712..101W}.
As each halo hosted consecutive cycles of stellar birth, H II region
formation and SN explosions \citep{2007ApJ...663..687Y}, gravity and
accretion flows congregated them into the first primitive galaxies
with halo merger time-scales of $\sim$ 20 Myr at $z \sim$ 20
\citep[e.g.][]{jgb08,get08,get10,wise12}. At the same time, the first
SNe enriched the IGM with metals and dust, triggering a transition
from Pop III to Pop II star formation
\citep[e.g.][]{mbh03,ss07,bsmith09,nho09,chiaki12}. How these two
processes determined the masses and formation rates of stars in
primeval galaxies at $z \sim$ 10 -- 15, and therefore their spectra
and luminosities, is unknown.

Stellar evolution models show that the final fates of the first stars
rested upon their masses at birth \citep{hw02}:\ $15 - 40\, {\rm
  M_\odot}$ Pop III stars died in core-collapse SNe \citep{jet09b} and
$40 - 140\, {\rm M_\odot}$ stars collapsed to black holes, perhaps
with violent pulsational mass loss prior to death \citep{wbh07}. Pop
III stars from 140 to 260~${\rm M_\odot}$ died as pair instability
(PI) SNe, extremely energetic thermonuclear explosions with energies
of up to 100 those of Type Ia and Type II SNe (Joggerst \& Whalen
2011; Chatzopoulos \& Wheeler 2012 extend this lower limit down to
$65\, {\rm M_\odot}$ for rotating stars). Some $40 - 60\, {\rm
  M_\odot}$ Pop III stars may have died as hypernovae, with energies
intermediate to those of core-collapse and PI SNe
\citep{Iwamoto2005}. Attempts have been made to indirectly constrain
the masses of the first stars by reconciling the cumulative
nucleosynthetic yield of their supernovae to the chemical abundances
found in ancient, dim metal-poor stars in the Galactic halo
\citep{bc05, fet05}, some of which may be contaminated with the ashes
of the first generation. For example, \citet{jet09b} recently
discovered that the average yields of $15 - 40\, {\rm M_\odot}$ Pop
III SNe agree well with the fossil abundances measured in $\sim130$
stars with metallicities $Z < 10^{-4} Z_{ \odot}$
\citep{Cayrel2004,Lai2008}, suggesting that low-mass primordial stars
may have synthesized most metals at high redshift. However, stellar
archaeology is still in its infancy because of small sample sizes,
systematics in the measurements of some elements, and because the
imprint of metals from first-generation stars on the second is not
well understood.

The direct detection of Pop III SNe may be the best prospect for
probing the earliest generation of stars in the near term. Primordial
supernovae may be 100,000 times brighter than their progenitors, or,
at slightly lower redshifts, the primitive galaxies in which they
reside. Their transience readily distinguishes them from early
galaxies, with which they otherwise overlap in colour--colour space.
Previous studies have investigated detection limits for PI SNe at $z
\sim 6$ \citep{sc05,tet12}, for 6 $< z <$ 15 \citep[][Whalen, Even et
  al. in preparation]{pan11,moriya12,2012arXiv1209.5459W}, and in very
approximate terms for $z \sim$ 30 \citep{2012ApJ...755...72H}. Whalen,
Fryer et al. (in preparation) and Whalen, Frey et al. (in preparation)
show that \textit{JWST} will detect PI SNe beyond $z \sim 30$ and that
the \textit{Wide Field Infrared Survey Telescope} (\textit{WFIRST})
and the \textit{Wide-field Imaging Surveyor for High-Redshift}
(\textit{WISH}) will detect them out to $z \sim$ 15 -- 20 in all-sky
NIR surveys. Unfortunately, it may be a decade before such
observations are possible, given that \textit{JWST} and
\textit{WFIRST} will not be in operation before 2018 and 2021,
respectively.

In the meantime, it may be possible to discover the first generation
of supernovae in cosmic backgrounds. Pair instability SNe deposit up
to half of their energy into the CMB via inverse Compton scattering at
$z \sim$ 20 \citep{ky05,2008ApJ...682...49W}. \citet{oh03} have found
that Pop III PI SNe may impose excess power on the CMB on small scales
via the Sunyayev-Zeldovich effect. Primordial SNe may also be manifest
as fluctuations in the NIR background because they are so much more
luminous than their host protogalaxies at high redshift.

In this paper, we show Pop III supernovae may also be revealed through
the detection of the radio emission from their remnants by both
current and future radio observatories such as the
eVLA\footnote{\texttt{www.aoc.nrao.edu/evla/}},
eMERLIN\footnote{\texttt{www.jb.man.ac.uk/research/rflabs/eMERLIN.html}},
MeerKAT\footnote{\texttt{public.ska.ac.za/meerkat}},
ASKAP\footnote{\texttt{www.atnf.csiro.au/projects/askap}} and the
Square Kilometre Array\footnote{\texttt{www.skatelescope.org}} (SKA).
A review of planned deep continuum radio surveys, including some that
would be capable of detecting the remnants discussed here, is provided
by \citet{2012arXiv1210.7521N}. \citet{2005ApJ...619..684I} previously
suggested the radio afterglows of high redshift hypernovae would
produce large fluxes while still relativistically expanding, however
this phase is brief so that the probability of discovery of such
systems in collapsing minihaloes is very low. The afterglows may be
more common in primitive galaxies with established star formation.

Previous studies of the observability of collapsing structures at
$z>10$ in the radio have centered on detecting the 21cm signal of
minihaloes during the cosmic dark ages
\citep[e.g.][]{2002ApJ...579....1F, 2002ApJ...572L.123I,
  2006ApJ...646..681S, 2011MNRAS.417.1480M}, not cosmic explosions at
first light. There are several advantages to searching for the first
supernovae through their radio synchrotron emission. Firstly, if only
certain types of explosions contribute a radio signal, this may be
used to place limits on the masses of their progenitors. Secondly,
supernova (and hence star formation) rates may possibly be determined
from the strength of the radio signals and the number counts of the
sources. Finally, depending on the properties of this signal, it may
be possible to pinpoint the redshift of the explosions.

In this paper we calculate the long term radio signal produced by the
remnants of core-collapse SNe, hypernovae and PI SNe to assess the
detectability of the remnants by current and future observatories. We
also determine if this signal may constrain the masses and formation
rates of Pop III stars in high redshift protogalaxies. In $\S \,
\ref{sec:sync}$ we review how Pop III SN explosions generate
synchrotron emission that could be directly detected by current and
planned radio telescopes. In $\S \, \ref{sec:obspred}$ we describe the
expected radio light curves and number counts of the Pop III SN radio
remnants. We summarise our conclusions in $\S \,
\ref{sec:conclusions}.$ Formulae are provided for synchrotron emission
and self-absorption in terms of the energy densities of the
relativistic electrons and the magnetic field in an appendix. The
range in relativistic gamma factors is discussed in a second appendix.

For numerical cosmological estimates, we adopt $\Omega_{\rm m}=0.27$,
$\Omega_{\rm v}=0.73$, $H_0=100h\,{\rm km}\,{\rm s}^{-1}\,{\rm
  Mpc}^{-1}$ with $h=0.70$, $\sigma_{\rm 8h^{-1}}=0.81$ and $n=0.96$
for the total mass and vacuum energy density parameters, the Hubble
constant, the linear density fluctuation amplitude on a scale of
$8\,h^{-1}$Mpc and the spectral index, respectively, consistent with
the best-fitting cosmological parameters for a flat universe as
constrained by the 5-yr {\it Wilkinson Microwave Anisotropy Probe}
({\it WMAP}) measurements of the Cosmic Microwave Background
\citep{2009ApJS..180..330K}.

\section{The synchrotron signature of supernovae in mini-haloes}
\label{sec:sync}

We begin by estimating the expected synchrotron flux from a supernova
remnant. For relativistic electrons with energy density $u_e$ in a
magnetic field of energy density $u_B$, the bolometric emissivity of
synchrotron radiation is
\begin{equation}
  \epsilon_{\rm bol}\sim u_e\tau_{\rm S}^{-1},
\label{eq:Psync_bol_ap}
\end{equation}
where $\tau_{\rm S}$ is the characteristic synchrotron cooling time
\begin{equation}
\tau_{\rm S}=\frac{3}{4}\frac{m_e c^2}{c \sigma_T \gamma_u u_B}.
\label{eq:tau_sync}
\end{equation}
for the most energetic electrons with kinetic energy
$(\gamma_u-1)m_ec^2$, assuming $\gamma_u>>1$. Here $\sigma_T$ is the
Thomson cross section for electron scattering and $m_e$ is the mass of
an electron.

Allowing for a fraction $f_e$ of the thermal energy to go into
relativistic electrons, so that $u_e=f_e u_{\rm th}$ for a thermal
energy density $u_{\rm th}$, and further assuming equipartition
$u_B\simeq u_e$\citep{1982ApJ...259..302C}, the bolometric synchrotron
power emitted by a single halo is
\begin{equation}
  P^{\rm sync}_{\rm bol}\simeq4\pi\gamma_u\frac{c \sigma_T}{m_e c^2}\int\,dr r^2
  f_e(r)^2 u_{\rm th}^2 F(p_e;\gamma_l,\gamma_u),
\label{eq:E_bol}
\end{equation}
where the function $F(p_e;\gamma_l,\gamma_u)$, described in
Appendix~\ref{app:Sync}, allows for a power-law energy distribution of
index $p_e$ for the relativistic electrons, and $\gamma_l$ and
$\gamma_u$ are the lower and upper ranges of the relativistic $\gamma$
factors for the electrons. The upper electron energies are generally
believed to be due to acceleration within the region of the shock,
although the precise mechanism is still unknown. Observations suggest
values as high as $\gamma_u\simeq10^8$ are achieved
\citep{1995Natur.378..255K}. For the supernova remnants examined here,
synchrotron cooling is sufficiently strong downstream of the shock
that $\gamma_u$ of a few hundred is more typical. A lower energy
cut-off $\gamma_l$ is imposed by energy losses arising from the
excitation of quantized plasma waves (plasmons). The computation of
the range in relativistic electron energies used here is discussed in
Appendix~\ref{app:gamrange}.

The electron energy spectrum will generally evolve downstream of the
shock due to synchrotron cooling, resulting in a steepening of the
synchrotron spectrum above a break frequency around 100~GHz. The
assumption of a constant $f_e$ and $p_e$ is therefore an approximation
and should only be regarded as effective values. The model none the
less qualitatively recovers the effects of the high frequency break,
as discussed in Appendix~\ref{app:gamrange}. The high frequency range
of the spectrum, however, should be regarded as uncertain by factors
of several.

We illustrate the expected flux using the computation for a $40\,{\rm
  M_\odot}$ hypernova in a $1.2\times10^7\,{\rm M_\odot}$ halo at
$z=17.3$ \citep{2008ApJ...682...49W}. The synchrotron power may be
estimated from the post-shock thermal energy density of the supernova
remnant, which exceeds a few ${\rm erg\,cm^{-3}}$ over a region around
a tenth of a parsec across during the first few years. For $f_e=0.01$
\citep[e.g.][]{1982ApJ...259..302C}, this corresponds to a bolometric
synchrotron power output of $P^{\rm sync}_{\rm
  bol}\simeq5\times10^{42}\,{\rm erg\,s^{-1}}$ for $\gamma_u=300$ and
$p_e=2$. The spectrum will extend to a lower characteristic cutoff
wavelength, assuming an isotropic magnetic field, of
$\lambda_c=(2\pi^{1/2}/3)(m_ec^2/e)\gamma_u^{-2}u_B^{-1/2}\simeq
2014\gamma_u^{-2}u_B^{-1/2}\,{\rm cm}\approx0.2\,{\rm cm}$, or upper
cutoff frequency $\nu_c\simeq130$~GHz, corresponding to a
characteristic specific power at $\nu_c$ of $2\times10^{31}\,{\rm
  erg\,s^{-1}\,Hz^{-1}}$. For a source at $z=17.3$, this corresponds
to an observed radio flux of $\sim10\,\mu$Jy. The flux will vary with
frequency as $\nu^{-(p_e-1)/2}=\nu^{-1/2}$.

\begin{figure}
\includegraphics[width=3.3in]{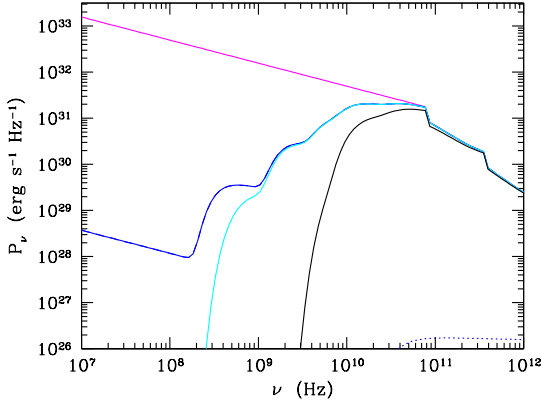}
\caption{Radio power from a $40\,{\rm M_\odot}$ hypernova in a
  $1.2\times10^7\,{\rm M_\odot}$ minihalo at $z=17.3$ (for $f_e=0.01$
  and $p_e=2$). Shown are total (synchrotron and thermal free-free)
  (solid lines) and thermal free-free (dotted line) powers. The upper
  curve (magenta) corresponds to a model with no attenuation. The
  middle curves allow for synchrotron self-absorption (blue) and
  free-free absorption as well (cyan). The lowest curve (black)
  further adds hypothesized plasma synchrotron attenuation effects.}
\label{fig:40Ms_halo3}
\end{figure}

The radio spectrum computed 4.7~yrs after the explosion is shown in
Fig.~\ref{fig:40Ms_halo3}. The thermal free-free radiation is shown as
well, but is generally much smaller than the synchrotron. The energy
density in relativistic electrons is sufficiently high that
synchrotron self-absorption (SSA) is significant at the lower
frequencies. Allowing for SSA severely attenuates the spectrum at
$\nu\lsim40$~GHz, as shown in Fig.~\ref{fig:40Ms_halo3}.

Adding free-free absorption degrades the spectrum further at low
frequencies. Free-free absorption was neglected from the fluid zone at
the shock front, where the gas is becoming ionized but has not yet
reached the post-shock temperature. Since the shock front is
unresolved, the substantial free-free absorption from it is likely
greatly over-estimated. For a shock front width on the order of the
Coulomb mean free path, the free-free absorption from the shock front
becomes negligible. It will therefore generally not be included
here. It is remarked, however, that if cool circumstellar gas mixes in
with the post-shock ionized gas, free-free absorption could be
non-negligible.  Such an effect is beyond the capacity of a
spherically symmetric code to reproduce, and would be difficult to
resolve in any case.

The Razin-Tsytovich plasma effect on the synchrotron radiation
mechanism may further degrade the escaping synchrotron radiation
\citep[e.g.][]{1968ARA&A...6..321S, 1986rpa..book.....R}. We
approximate the effect by cutting off the production of synchrotron
radiation by the factor $\exp(-\nu_{\rm RT}/\nu)$, where
\begin{equation}
\nu_{\rm RT} = \gamma_u\nu_p\left(1+\gamma_u\frac{\nu_p}{\nu_c}\right),
\label{eq:nuRT}
\end{equation}
and $\nu_p$ is the electron plasma frequency. This strongly cuts off
most of the radiation below $\sim0.1\nu_c$, as shown in
Fig.~\ref{fig:40Ms_halo3}. As the occurrence of the hypothesized
effect is unclear, we conservatively neglect it further here except as
noted, but remark that the detection of such greatly diminished
low-frequency emission would support its reality.

Radio observations of supernova remnants do not strongly constrain
$f_e$, although values of $f_e\sim0.001-0.2$ are indicated
\citep{1986ApJ...301..790W, 1998ApJ...499..810C,
  2006ApJ...651.1005S}. Assuming equipartition, the power at
$\nu=\nu_c$ is proportional to $f_e^{3/2}$. It is worth noting that
for $f_e<<1$, an equipartition magnetic field will contribute
negligibly to the pressure forces acting on the post-shock gas, so
that the hydrodynamical models used should not be much affected.

\section{Observational predictions}
\label{sec:obspred}

Deep radio surveys are currently able to reach the few $\mu$Jy
level. The Very Large Array has achieved an {\it rms} noise level of
$1.5\,\mu$Jy at 8.4~GHz \citep{2002AJ....123.2402F} and $2.7\,\mu$Jy
at 1.4~GHz \citep{2008AJ....136.1889O}. Similar levels have been
reached using eMERLIN \citep{2005MNRAS.358.1159M}. The Square
Kilometre Array is projected to be a factor 100--1000 more sensitive.

\begin{figure}
\begin{center}
\leavevmode
\epsfig{file=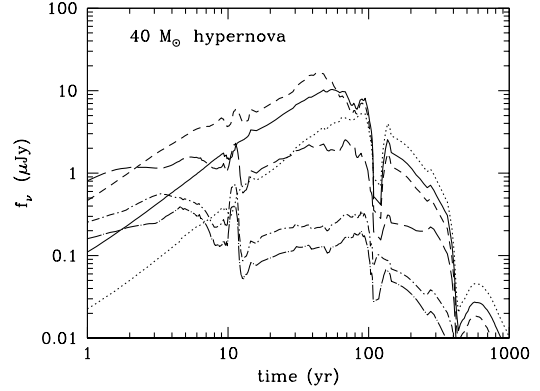, height = 8cm}
\epsfig{file=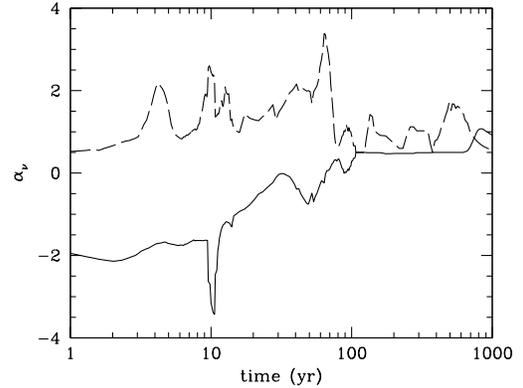, height = 8cm}
\end{center}
\caption{Top panel:\ Radio light curves for a $40\,{\rm M_\odot}$
  hypernova in a $1.2\times10^7\,{\rm M_\odot}$ minihalo at $z=17.3$
  (for $p_e=2$ and $f_e=0.01$). The curves correspond to the
  bands:\ 0.5 (dotted), 1.4 (solid), 3 (short-dashed), 10
  (long-dashed), 25 (dot short-dashed) and 35 (dot long-dashed)
  GHz. Bottom panel:\ The corresponding spectral indices at 1.4
  (solid) and 10 (long-dashed) GHz.}
\label{fig:40Ms_lightcurves}
\end{figure}

\begin{figure}
\begin{center}
\leavevmode
\epsfig{file=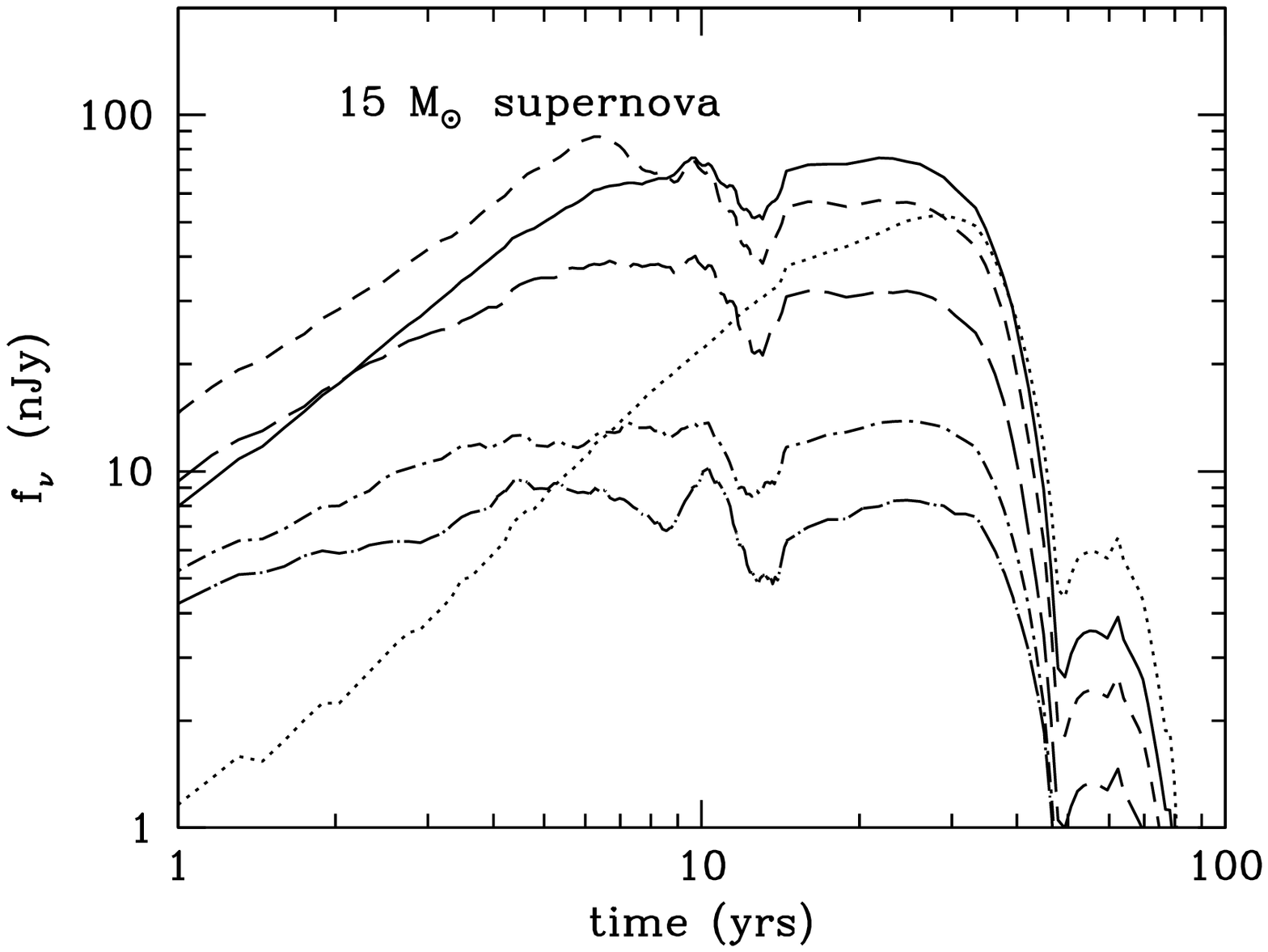, height = 8cm}
\epsfig{file=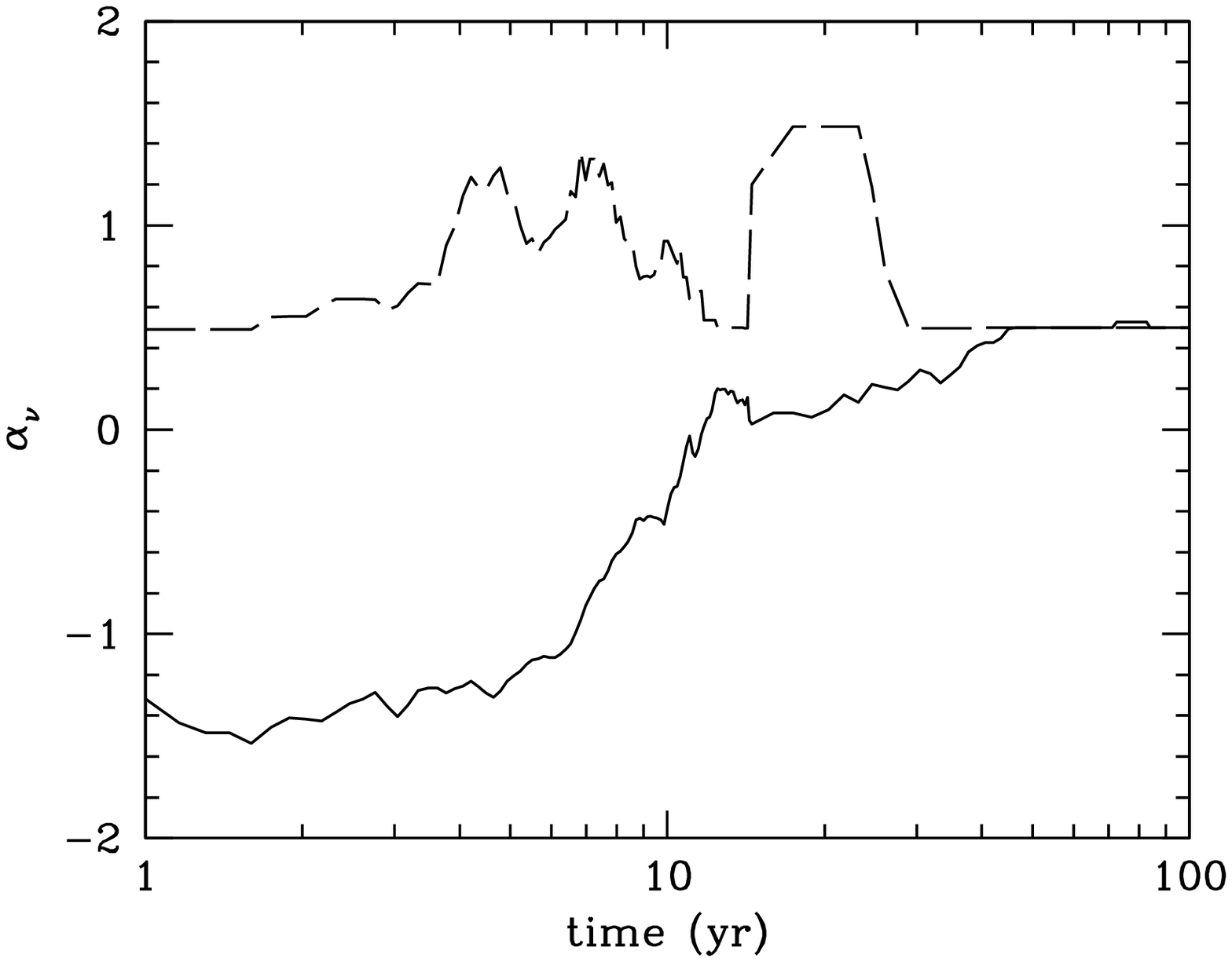, height = 8cm}
\end{center}
\caption{Top panel:\ Radio light curves for a $15\,{\rm M_\odot}$ Type
  II supernova in a $1.2\times10^7\,{\rm M_\odot}$ minihalo at
  $z=17.3$ (for $p_e=2$ and $f_e=0.01$). The curves correspond to the
  bands:\ 0.5 (dotted), 1.4 (solid), 3 (short-dashed), 10
  (long-dashed), 25 (dot short-dashed) and 35 (dot long-dashed)
  GHz. Bottom panel:\ The corresponding spectral indices at 1.4
  (solid) and 10 (long-dashed) GHz.}
\label{fig:15Ms_lightcurves}
\end{figure}

In Figs~\ref{fig:40Ms_lightcurves} and \ref{fig:15Ms_lightcurves}, we
show the light curves for a $40\,{\rm M_\odot}$ hypernova and
$15\,{\rm M_\odot}$ Type II supernova, respectively, in a
$1.2\times10^7\,{\rm M_\odot}$ minihalo at $z=17.3$ in the observed
frame for selected bands planned for the SKA. The bands lie within the
eVLA frequency range. The eMERLIN L-band includes 1.4~GHz, while the
C-band flux light curve would lie between those for the 3 and 10~GHz
bands. For the $40\,{\rm M_\odot}$ hypernova, time-dilation produces a
gradual rise in the observed flux $f_\nu$ extended over a 100~yr
time-scale with an even slower decline. A flux exceeding $1\,\mu$Jy is
expected to persist over a span of 300~yrs, with a peak flux of
$\sim10\,\mu$Jy. Most notable is the relatively late rise of the flux
in the 500~MHz band; it eventually overtakes the higher frequency
fluxes for a period of 150~yrs while above $1\,\mu$Jy.

Sizable fluctuations in the flux over time are found. These arise
because the dominant emitting region lies just downstream of the shock
front with a width governed by synchrotron and plasmon excitation
losses, as described in Appendix~\ref{app:gamrange}. The width
increases with decreasing gas density and magnetic field strength and
decreases with decreasing shock velocity. As the shock slows and
weakens, these factors work against each other, producing a
fluctuating volume of emitting gas. For intermittent periods, the
energy losses deplete the relativistic electron population except very
near the shock front.

Since radio observations are often quantified by the spectral index
$\alpha_\nu=-d\log f_\nu/d\log \nu$, the spectral indices at two
representative SKA bands are shown in the lower panel. The changing
structure of the post-shock gas produces varying spectral indices in
the period leading up to the peak emission. The low frequency
attenuation results in $\alpha_\nu<0$ at frequencies below 3~GHz. At
late times, after the peak emission, the emission is dominated by a
restricted region downstream of the shock front for which
$\gamma_u>1000$, and the spectral index at the lower frequencies
settles to $\alpha_\nu=0.5$, the value expected for a power-law
relativistic electron energy distribution with $p_e=2$.

Similar trends are found for the $15\,{\rm M_\odot}$ supernova,
although over the much shorter time-scale of decades and with flux
values lower by a factor of $\sim100$. We also computed the emission
following a $260\,{\rm M_\odot}$ pair instability supernova, but found
it so efficiently sweeps through the minihalo that the resulting
observed radio flux is at most $\sim0.1$~nJy. Similarly, we found the
observed radio fluxes from the remnants produced in a
$2\times10^6\,{\rm M_\odot}$ halo by all three supernova progenitor
masses fell well below 1~nJy.

\begin{figure}
\begin{center}
\leavevmode
\epsfig{file=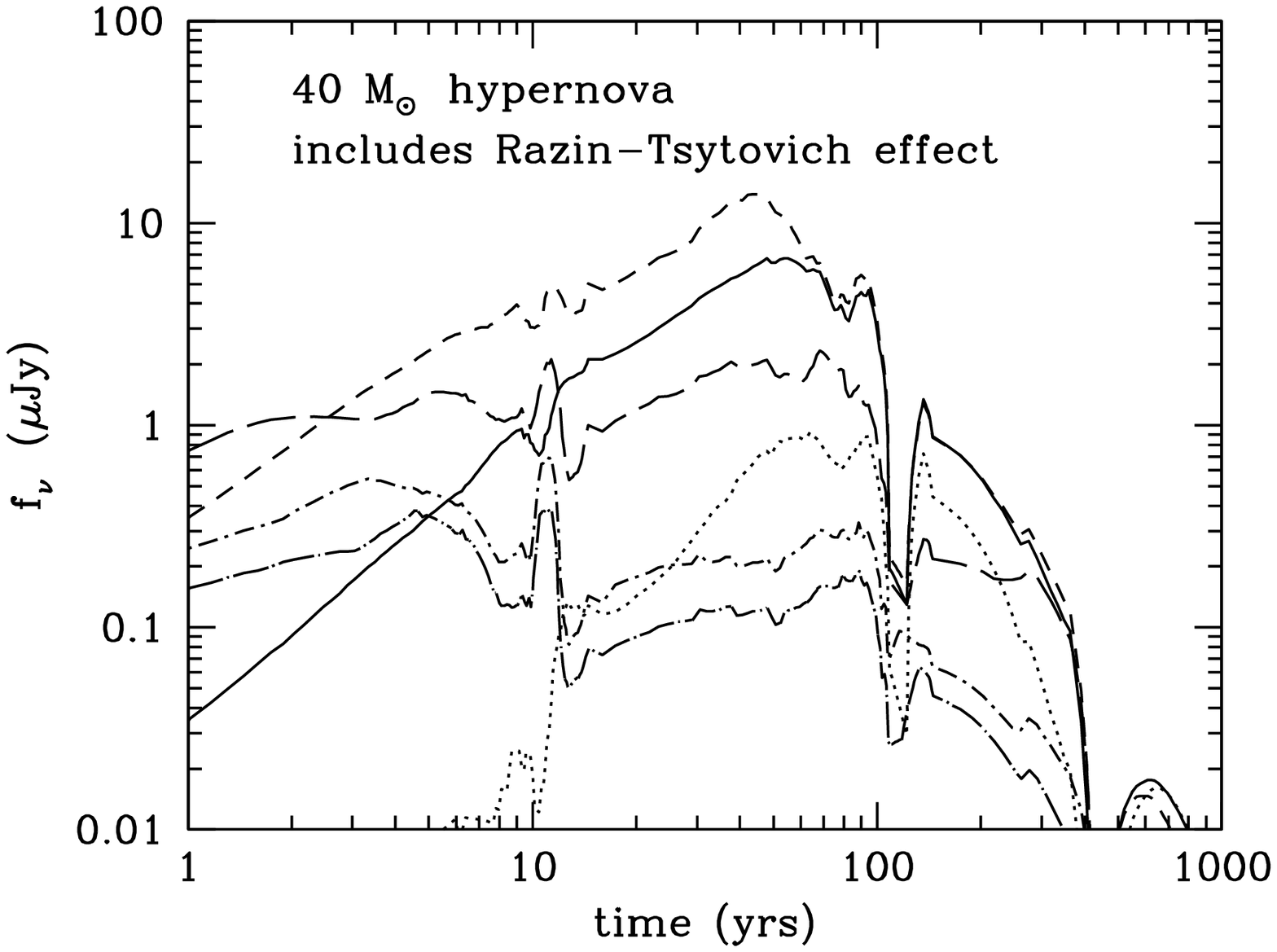, height = 8cm}
\epsfig{file=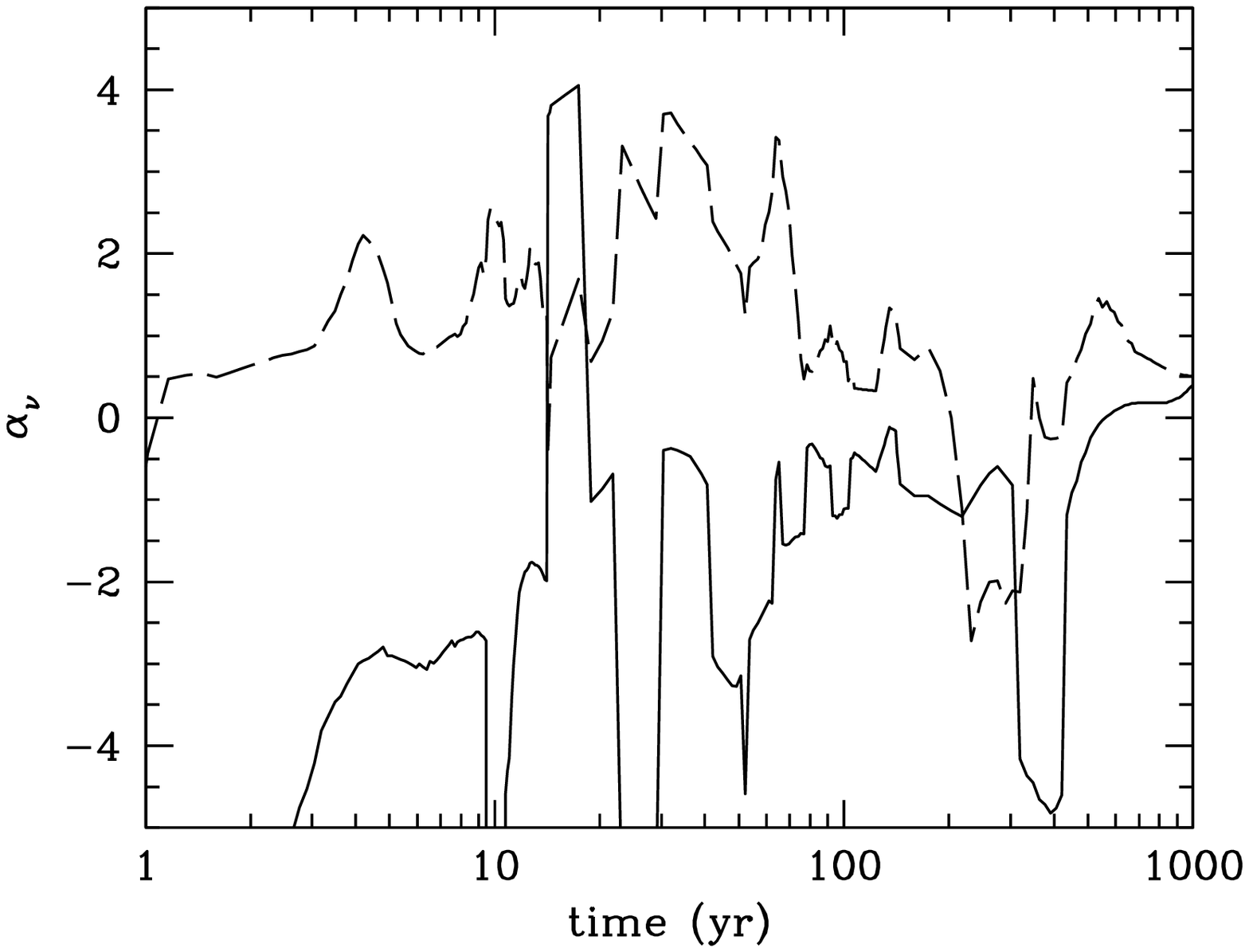, height = 8cm}
\end{center}
\caption{Top panel:\ Radio light curves for a $40\,{\rm M_\odot}$
  hypernova in a $1.2\times10^7\,{\rm M_\odot}$ minihalo at $z=17.3$
  (for $p_e=2$ and $f_e=0.01$). The Razin-Tsytovich effect is
  included. The curves correspond to the bands:\ 0.5 (dotted), 1.4
  (solid), 3 (short-dashed), 10 (long-dashed), 25 (dot short-dashed)
  and 35 (dot long-dashed) GHz. Bottom panel:\ The corresponding
  spectral indices at 1.4 (solid) and 10 (long-dashed) GHz.}
\label{fig:40Ms_RT_lightcurves}
\end{figure}

Allowing for the Razin-Tsytovich effect much reduces the flux in the
500~MHz band, as shown in Fig.~\ref{fig:40Ms_RT_lightcurves}. While it
continues to rise with time, it lies about 1~dex below the flux at
1.4~GHz for 50~yrs, and never overtakes it. Comparison with
Fig.~\ref{fig:40Ms_lightcurves} suggests the measured flux in the
500~MHz band may be used to detect the presence of the Razin-Tsytovich
effect. The spectrum at 500~MHz is very steep, with $\alpha_\nu<-2.5$
during the first 100~yrs.

The number of sources visible per year is substantial. We estimate
this from the halo collapse rate using the halo fitting function of
\citet{2007MNRAS.374....2R}, adapted to the 5~yr {\it WMAP}
cosmological parameters for a flat universe
\citep{2009ApJS..180..330K}. The observed formation rate of haloes
within a solid angle $\delta\Omega$ with a comoving number density
$n_h^{\rm com}(z)$ for halo masses exceeding $M_h$, integrated over
$z_1 < z < z_2$, is then
\begin{eqnarray}
  {\dot N}_h^{\rm obs}(>M_h) &=& \int_{z_1}^{z_2} dz\,{\dot n}_h^{\rm com}(1+z)^3\frac{dV^{\rm prop}}{dz}\frac{1}{1+z}\nonumber\\
  &=&-\int_{z_1}^{z_2}dz\, \frac{dn_h^{\rm com}}{dz}c(a_0r)^2\delta\Omega\nonumber\\
  &=&\left[n_h^{\rm com}c(a_0r)^2\right]_{z_2}^{z_1}\delta\Omega\nonumber\\
  &&+2\left(\frac{c}{H_0}\right)\int_{z_1}^{z_2}dz\,n_h^{\rm
    com}c\frac{a_0r}{[\Omega_{\rm m}(1+z)^3+\Omega_{\rm v}]^{1/2}}\delta\Omega\nonumber\\
  &\simeq&n_h^{\rm com}(z_1)c[a_0r(z_1)]^2\delta\Omega,
\label{eq:dotNh}
\end{eqnarray}
where $a_0r(z) = (c/H_0)\int_0^z dz\, [\Omega_{\rm
  m}(1+z)^3+\Omega_{\rm v}]^{-1/2}$ is the angular size distance for a
flat universe with present day mass and vacuum energy density
parameters $\Omega_{\rm m}$ and $\Omega_{\rm v}$, respectively, and
Hubble constant $H_0$. For $z_1>3$,
$(H_0/c)a_0r(z_1)\simeq1.5+1.9[1-2/(1+z_1)^{1/2}]$ to better than 2
percent accuracy. The approximation ${\dot n}_h^{\rm com}=(dn_h^{\rm
  com}/dz)(dz/dt)$ was made, although this could be modified by
allowing for merger histories. In the final line, the integral in the
line previous was neglected as was the halo density at $z=z_2$. A
factor $1/(1+z)$ has been included to account for time-dilation.

We find formation rates for haloes $M>10^7\, h^{-1}\,{\rm M_\odot}$ and
$z>20$ of $0.0016\,{\rm deg^{-2}\,yr^{-1}}$, and $0.20\,{\rm
  deg^{-2}\,yr^{-1}}$ at $z>10$. These amount to about 70 to 8000
collapsing haloes per year over the sky. The rates are comparable to
recent predictions for the total production rate of Pop~III pair
instability supernovae, with progenitor masses between
$140-260\,{\rm M_\odot}$, based on cosmological simulations
\citep[e.g.][]{2012ApJ...755...72H, 2012arXiv1206.5824J}, assuming one
PI SN per minihalo formed. On the other hand, small scale simulations
suggest fragmentation may prevent the formation of Pop~III stars
sufficiently massive to form a PI SN \citep[e.g.][]{stacy10}, although
subsequent mergers of the protostars into more massive ones cannot be
ruled out. The star formation may also be stochastic given the low
number of massive stars formed, with hypernovae forming before a more
massive PI SN progenitor is created.

The effect of multiple supernovae on the gas in small haloes is
unknown. The heat input of a supernova is sufficient to unbind the IGM
of a minihalo with mass below $\sim10^7\,{\rm M_\odot}$, while, once the gas
cools, it will fall back in a more massive system
\citep{2008ApJ...682...49W, 2011MNRAS.417.1480M}. The remaining
massive stars will further replenish the IGM through winds, so a
gaseous environment may continue to persist on which further
supernovae may impact, producing synchrotron emitting remnants. If
each halo gives rise to at least one supernova with progenitor mass
exceeding $40\, {\rm M_\odot}$, its remnant should then be detectable by eVLA
or eMERLIN. If each produces one with a progenitor mass exceeding $15\,
{\rm M_\odot}$, SKA should be able to detect the remnant.

An estimate of the expected number counts of radio emitting remnants
requires modelling the remnants over a broad redshift range. In the
absence of simulations covering a wide range, we use those discussed
above, presuming the behaviour of the gas is not very sensitive to
redshift. This is a coarse approximation, as the gas content of the
haloes is expected to be sensitive to the redshift of their
formation. On the other hand, since much of the gas into which the
supernova explodes is produced by stellar winds, the environment of
the progenitor may not be very redshift dependent. An improved
estimate would require a better understanding of star formation within
primordial haloes.

\begin{figure}
\begin{center}
\leavevmode
\epsfig{file=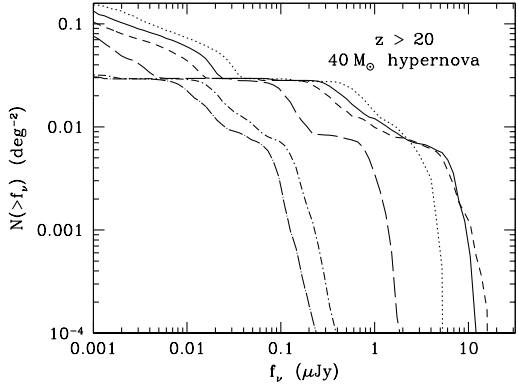, height = 8cm}
\epsfig{file=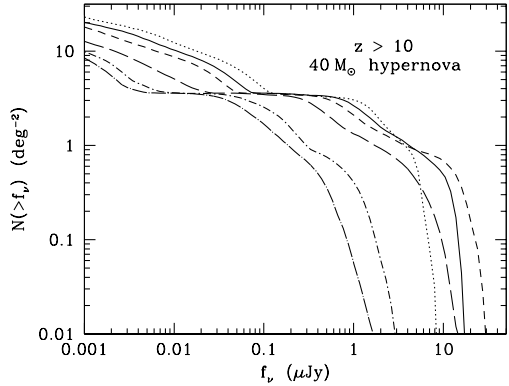, height = 8cm}
\end{center}
\caption{Number counts of radio remnants above a given flux, based on
  radio light curves for a $40\,{\rm M_\odot}$ hypernova in
  $1.2\times10^7\,{\rm M_\odot}$ minihaloes (for $p_e=2$ and
  $f_e=0.01$) at $z>20$ (upper panel) and $z>10$ (lower panel). The
  observed counts are shown for bands:\ 0.5 (dotted), 1.4 (solid), 3
  (short-dashed), 10 (long-dashed), 25 (dot short-dashed) and 35 (dot
  long-dashed) GHz.}
\label{fig:40Ms_counts}
\end{figure}

\begin{figure}
\begin{center}
\leavevmode
\epsfig{file=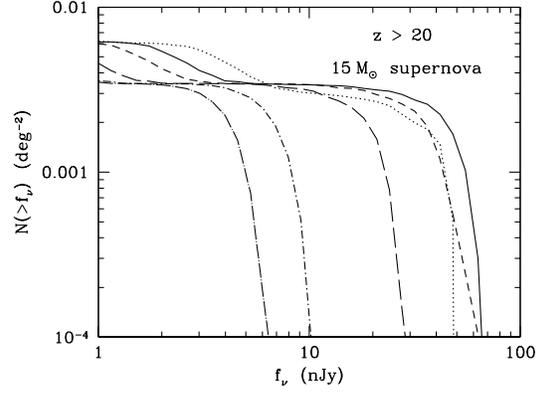, height = 8cm}
\epsfig{file=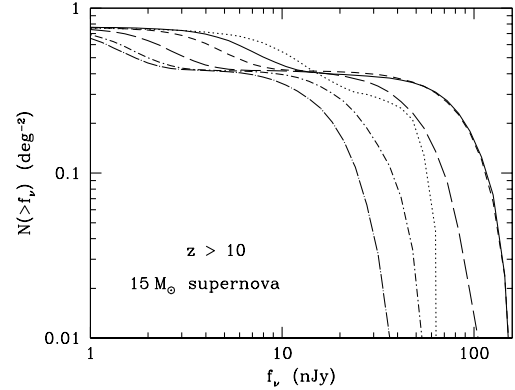, height = 8cm}
\end{center}
\caption{Number counts of radio remnants above a given flux, based on
  radio light curves for a $15\,{\rm M_\odot}$ Type II supernova in
  $1.2\times10^7\,{\rm M_\odot}$ minihaloes (for $p_e=2$ and
  $f_e=0.01$) at $z>20$ (upper panel) and $z>10$ (lower panel). The
  observed counts are shown for bands:\ 0.5 (dotted), 1.4 (solid), 3
  (short-dashed), 10 (long-dashed), 25 (dot short-dashed) and 35 (dot
  long-dashed) GHz.}
\label{fig:15Ms_counts}
\end{figure}

The number of sources with observed flux exceeding $f_\nu$ is given by
modifying equation~(\ref{eq:dotNh}) to account for the duration
$\delta t^{\rm eff}_{f_\nu}(z)$ the remnant is visible with a flux
exceeding $f_\nu$:
\begin{equation}
N(>f_\nu)=-\int_{z_1}^{z_2}dz\, \frac{dn_h^{\rm com}}{dz}c(a_0r)^2
\delta t^{\rm eff}_{f_\nu}(z) \delta\Omega.
\label{eq:dNdS}
\end{equation}
The resulting number counts for haloes collapsing at $z>20$ are shown
in Fig.~\ref{fig:40Ms_counts}, based on one $40\,{\rm M_\odot}$
hypernova per collapsing halo with a mass exceeding
$10^7\,h^{-1}\,{\rm M_\odot}$. In the lower frequency bands, the
counts decline above $0.3\,\mu$Jy but are still substantial to
$\sim10\,\mu$Jy. A source brighter than $1\,\mu$Jy should be visible
in the range $0.5<\nu<3$GHz every 100 square degrees.

Few young sources are found below $1\,\mu$Jy. This is because at a
rest-frame age of about 21~yrs, the width of the emitting region
diminishes to about $10^{15}\,{\rm cm}$, resulting in very little
emission (see Appendix~\ref{app:gamrange}), and a plateau in the
number counts. By an age of 32~yrs, as the shock further weakens, the
emission zone broadens to about $3\times10^{16}\,{\rm cm}$, and the
flux somewhat recovers (as is visible in
Fig.~\ref{fig:40Ms_lightcurves} after 400~yrs in the observed
frame). This produces a rise in the number counts at very low flux
levels, giving about one source brighter than 1~nJy per 10 square
degrees in the low frequency bands.

At the earliest epochs of star formation, although the remnants would
still have observed peak fluxes brighter than $1\,\mu$Jy, the
frequency of their occurrence becomes very small. We find only one
remnant at $z>30$ brighter than 1~nJy is expected over the entire sky.

\citet{2012arXiv1206.5824J} suggest Pop~III star formation may persist
to $z<10$. This substantially boosts the counts to a few per square
degree brighter than $1\,\mu$Jy, and more than 10 per square degree
brighter than 1~nJy. The sources may be identified by the weakness of
the 25 and 35~GHz fluxes compared with the lower frequency fluxes, a
signature of strong synchrotron cooling, although the fluxes at these
high frequencies are difficult to predict because of the effects of
energy loss on the energy distribution of the relativistic electrons
(see Appendix~\ref{app:gamrange}). More definitive would be a survey
campaign with repeat multiband observations extended over several
years to trace the light curves, especially during the early rising
phase.

The corresponding number counts based on one $15\,{\rm M_\odot}$ Type
II supernova per collapsing halo are shown in
Fig.~\ref{fig:15Ms_counts}. While the fluxes are considerably weaker
than for the more energetic hypernovae, the numbers are lower also as
a result of their shorter durations. To detect the remnants at $z>20$,
100--200 square degree fields would be required to detect remnants as
weak as 1~nJy. The counts for $z>10$ are considerably higher,
requiring now 1--2 square degree fields to detect the supernova
remnants at the 1~nJy level.

We have also examined the radio absorption signature of the remnants
against a bright background radio source. While synchrotron absorption
will produce a strong absorption feature at rest frame frequencies
below 20~GHz, it persists for less than $\sim100$~yr. The number of
detectable features along a line of sight will be
\begin{equation}
\frac{dN(>\tau_\nu)}{dz}=n_h^{\rm com}\pi(r_0^{\rm com})^2(1+z)
c\delta t^{\rm eff}_{\tau_\nu}(z),
\label{eq:dNdz}
\end{equation}
where $r_0^{\rm com}$ is the comoving radius of the remnant out to
which the line-of-sight absorption optical depth exceeds $\tau_\nu$
and $\delta t^{\rm eff}_{\tau_\nu}(z)$ is the effective duration of
the feature. For a characteristic comoving radius of $\sim1$~pc,
$dN/dz\sim3\times10^{-14}$ at $z=10$, so that the feature would be
undiscoverable. At late times, at $t\sim0.8$~Myr, the outflowing gas
cools sufficiently to produce a 21cm absorption feature that would be
measurable by SKA. We find a characteristic 21cm absorption doublet
arising from the cooling shell would form along lines of sight within
$\sim5$~pc (proper) of the centre of the minihalo. The feature would
survive about 1~Myr before transforming into a singlet absorption
feature indistinguishable from those expected from minihaloes. This
corresponds to $dN/dz \sim 10^{-6}$, again too small to be discovered
as there would still not be nearly an adequate number of bright
background radio sources to have a fair chance of seeing even one
feature.

\section{Conclusions}
\label{sec:conclusions}

We estimate the radio signatures of Pop~III supernovae in
$\sim10^7\,{\rm M_\odot}$ minihaloes, sufficiently massive to retain
their baryons and form supernova remnants, based on hydrodynamical
computations of $15\,{\rm M_\odot}$ Type II supernovae, $40\,{\rm
  M_\odot}$ hypernovae and $260\,{\rm M_\odot}$ pair instability
supernovae at $z=17.3$ within the haloes. We model the synchrotron
emission and absorption assuming a power-law distribution of
relativistic electron energies with a total energy density
proportional to the thermal energy of the ionized gas, and
equipartition between the relativistic electron and magnetic field
energies. Allowing for a relativistic electron component with total
energy one to ten percent of the gas thermal energy is sufficient for
a hypernova to produce observable fluxes at $0.5-10$~GHz exceeding
$1\,\mu$Jy and for the less massive Type II supernova to produce
observable fluxes exceeding 10~nJy at $0.5-25$~GHz.

The PI SN expels much of the halo gas, leaving behind a supernova
remnant that produces fluxes of at most 0.1~nJy. A halo mass exceeding
$10^8\,h^{-1}\,{\rm M_\odot}$ would be sufficient to retain the
baryons following a PI SN. Although we do not currently have
computations of PI SN in such massive haloes, we may estimate the halo
formation rates:\ $0.011\,{\rm deg}^{-2}\,{\rm yr}^{-1}$ if forming at
$z>10$, and $9.1\times10^{-6}\,{\rm deg}^{-2}\,{\rm yr}^{-1}$ for
those forming at $z>20$. These are about a factor 20--200 smaller than
the formation rates of minihaloes with masses exceeding
$10^7\,h^{-1}\,{\rm M_\odot}$, but would still produce a detectable
yield of PI SNe remnants over the sky provided their synchrotron
fluxes were sufficiently high.

Hypernovae at $z=17.3$ produce observed light curves that rise
gradually over a period of 100~yrs, and decline slowly over the
subsequent 200--300~yrs. By contrast, the Type II curves have rise
times of 5--10~yrs at frequencies above 1.4~GHz, and decline abruptly
after another 30~yrs. The 500~MHz flux lags the higher frequency bands
for both supernovae and dominates the emission shortly after it peaks
as the synchrotron spectrum reddens due to the weakening of the
supernova shock. Light curves computed with and without the
hypothesized Razin-Tsytovich effect show that the observed 500~MHz
flux is substantially diminished by the effect, suggesting it may
provide a useful means of testing for the presence of the effect.

The light curves for both the hypernova and Type II supernova remnants
in collapsing minihaloes may be distinguished from their galactic
counterparts by the weakness of their high frequency fluxes. The flux
decreases below the power-law prediction at rest-frame frequencies
above 100~GHz as a consequence of intense synchrotron cooling
downstream of the shock front in the high gas density environments of
the minihaloes, curtailing the maximum synchrotron
frequency. Predictions of the precise shape of the high frequency
spectrum, however, are complicated by the transport of relativistic
electrons in the presence of strong synchrotron and plasmon excitation
energy losses and the unknown magnetic field distribution.

Estimating the number counts of the supernovae from the minihalo
formation rate at $z>20$, we find a supernova formation rate of about
one per 600 square degrees per year. If the supernovae are able to
form down to $z=10$, the rate increases to one per 5 square degrees
per year. On average, one $z>20$ hypernova radio remnant with flux
exceeding $1\,\mu$Jy should be detectable at any given time per 100
square degree field, and one with a flux above 1~nJy per 10 square
degree field. If Pop III star formation persists to $z<10$, then a few
hypernova remnants brighter than $1\,\mu$Jy would be visible per
square degree, and more than ten per square degree brighter than
1~nJy.

Because of their smaller explosion energies, Type II supernovae will
produce weaker radio remnants with shorter durations. Fields of
100--200 square degrees would be required to detect remnants forming
at $z>20$ with fluxes above 1~nJy. The numbers are much improved for
supernovae forming down to $z<10$, requiring instead 1--2 square
degree fields to detect their radio remnants.

Pop~III supernova remnants with radio fluxes exceeding $1\,\mu$Jy
should be visible in radio surveys using the eVLA or eMERLIN, or
possibly the MIGHTEE survey conducted with the SKA pathfinder
MeerKAT. Weaker few nJy sources would be detectable by the SKA in
large numbers. An observing campaign extended over several years would
be able to follow the light curves, yielding invaluable data on the
formation of the first stars in the Universe and their impact on their
local environments.

The readily distinguishable differences between the radio light curves
of hypernova and Type II supernova remnants provide the possibility of
exploiting the radio emission to infer the initial mass function of
the first stars. The frequency with which these explosions are
detected over the sky will also constrain Pop III star formation rates
through cosmic time along with all-sky NIR surveys by \textit{WFIRST}
and \textit{WISH} and deep-field surveys by \textit{JWST}. Their
discovery in the radio will permit followup by deep-field observations
of the host galaxies by \textit{JWST} and the TMT. Surveys detecting
Pop III radio remnants at $10 < z < 15$ may also constrain the
multiplicity of their occurrence within individual primeval galaxies,
and even the large-scale distribution of the galaxies themselves. The
detection of primordial supernova remnants promises to be one of the
most spectacular discoveries in high-redshift radio astronomy over the
coming decade.

\begin{appendix}
\section{Synchrotron emission and absorption}
\label{app:Sync}

\begin{figure}
\includegraphics[width=3.3in]{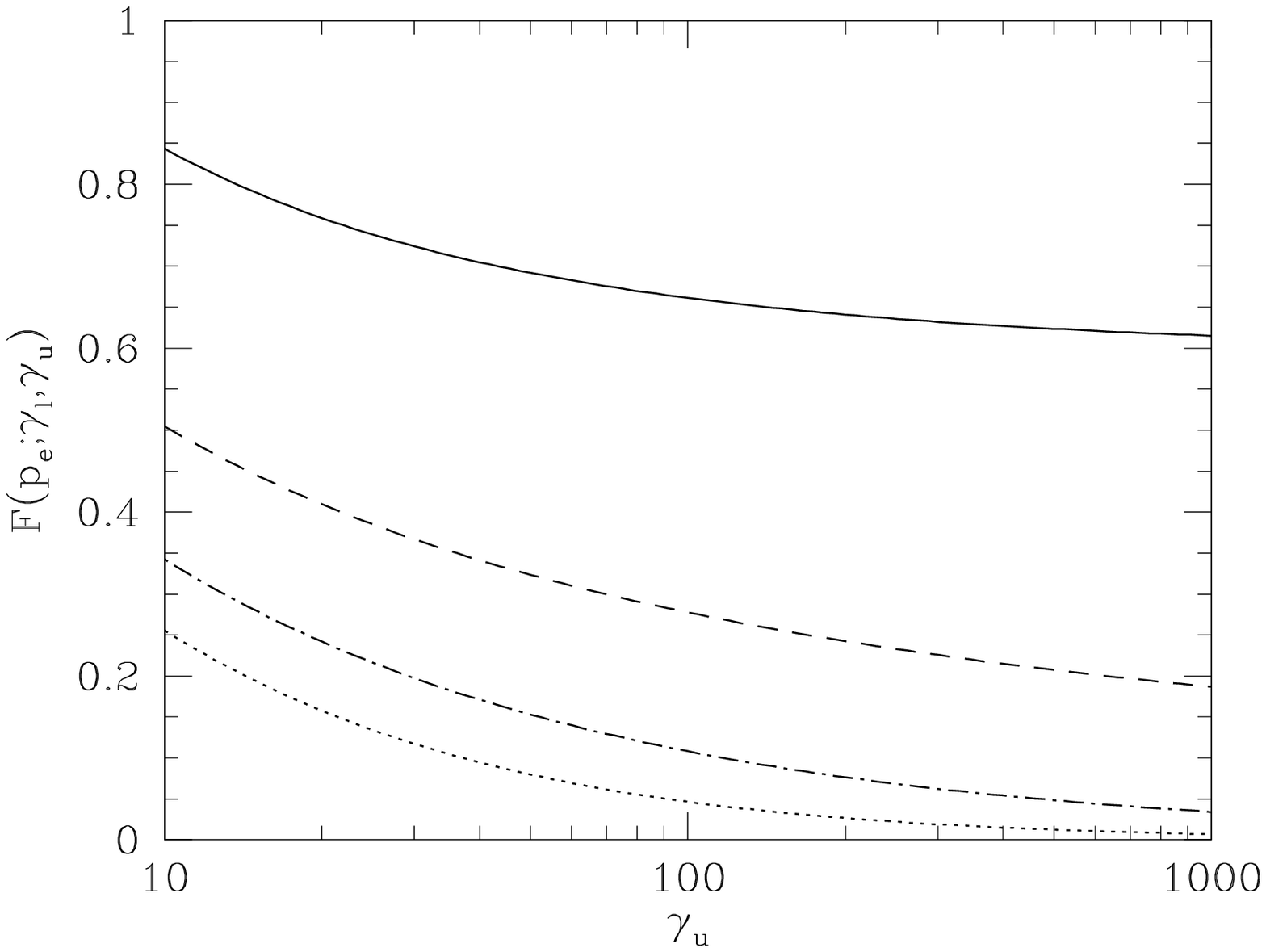}
\caption{Synchrotron function $F(p_e;\gamma_l,\gamma_u)$ for
  relativistic electron power-law energy distribution over
  $\gamma_l<\gamma<\gamma_u$ with index $p_e$. Shown for $p_e=1.5$
  (solid line), $p_e = 2.0$ (dashed line), $p_e=2.5$ (dot-dashed line)
  and $p_e=3.0$ (dotted line). (A value $\gamma_l=1$ is adopted.)}
\label{fig:Fp}
\end{figure}

The synchrotron emissivity for a power-law relativistic electron
energy distribution $dN/d\gamma\sim\gamma^{-p_e}$ extending over the
$\gamma$ factor range $\gamma_l<\gamma<\gamma_u$ may be expressed
concisely as
\begin{equation}
\epsilon_\nu=\begin{cases}\frac{3-p_e}{2}\frac{\epsilon_{\rm bol}}{\nu_c}
\left(\frac{\nu}{\nu_c}\right)^{-\frac{p_e-1}{2}}
\qquad ;\frac{1}{3}<p_e<3 \\
\frac{1}{\log(\nu_c/\nu_{\rm min})}\frac{\epsilon_{\rm bol}}{\nu_c}
\left(\frac{\nu}{\nu_c}\right)^{-1}
\qquad ;p_e=3,\end{cases}
\label{eq:epsnu}
\end{equation}
where $\nu_c$ is the synchrotron upper cutoff frequency
\begin{equation}
\nu_c=\frac{3}{4\pi}\frac{\gamma_u^2 eB\sin\alpha}{m_ec}
\label{eq:nuc}
\end{equation}
for pitch angle $\alpha$, and $\epsilon_{\rm bol}$ is the bolometric
emissivity
\begin{equation}
  \epsilon_{\rm bol}= \frac{3}{4}u_e\tau_{\rm S}^{-1}F(p_e;\gamma_l,\gamma_u).
\end{equation}
Here, $u_B$ is the energy density of the magnetic field, the
synchrotron cooling time $\tau_{\rm S}$ is given by
equation~(\ref{eq:tau_sync}) and the function
$F(p_e;\gamma_l,\gamma_u)$, shown in Fig.~\ref{fig:Fp}, is
\begin{eqnarray}
F(p_e;\gamma_l,\gamma_u) &=& 3^{1/2}\frac{9}{4\pi}\frac{2^{(p_e-1)/2}}{p_e+1}
\Gamma\left(\frac{p_e}{4} + \frac{19}{12}\right)
\Gamma\left(\frac{p_e}{4} - \frac{1}{12}\right)\nonumber\\
&&\times\begin{cases}
\frac{2-p_e}{3-p_e}\frac{1-(\gamma_l/\gamma_u)^{3-p_e}}
{1-(\gamma_l/\gamma_u)^{2-p_e}}\qquad ;p_e\neq2, p_e\neq3\\
\frac{1-\gamma_l/\gamma_u}{\log(\gamma_u/\gamma_l)}\qquad\qquad ;p_e=2\\
\frac{\log(\gamma_u/\gamma_l)}{\gamma_u/\gamma_l-1}\qquad\qquad ;p_e=3.
\end{cases}
\label{eq:Fp}
\end{eqnarray}

The inverse attenuation length due to synchrotron self-absorption may be
expressed as
\begin{equation}
\alpha_\nu=r_0^{-1}\left(\frac{1}{\nu_c\tau_{\rm se}}\right)
\left(\frac{\nu_c}{\nu}\right)^{(p_e+4)/2}G(p_e;\gamma_l,\gamma_u),
\label{eq:alphanu}
\end{equation}
where $r_0$ is the classical electron radius, $\tau_{\rm se}=\gamma_u
m_e c^2/c \sigma_T u_e$ is the characteristic electron scattering
time, and
\begin{eqnarray}
G(p_e;\gamma_l,\gamma_u)&=&\frac{3^{1/2}}{16\pi}\frac{2^{p_e/2}}{\gamma_u^3}
\Gamma\left(\frac{p_e}{4} + \frac{1}{6}\right)
\Gamma\left(\frac{p_e}{4} + \frac{11}{6}\right)\nonumber\\
&&\times\begin{cases}\frac{2-p_e}{1-(\gamma_l/\gamma_u)^{2-p_e}}\qquad ;p_e\neq2\\
\frac{1}{\log(\gamma_u/\gamma_l)}\qquad ;p_e=2.
\end{cases}
\label{eq:Gp}
\end{eqnarray}

\section{Range in relativistic electron energies}
\label{app:gamrange}
The dominant energy loss processes of relativistic electrons in the
supernova remnants examined are synchrotron cooling, bremsstrahlung
losses through scattering off the background thermal distribution of
ions, and the excitation of plasmon waves. We discuss each of these in
turn, using the rate estimates from \citet{1975ApJ...196..689G}.

\begin{figure}
\includegraphics[width=3.3in]{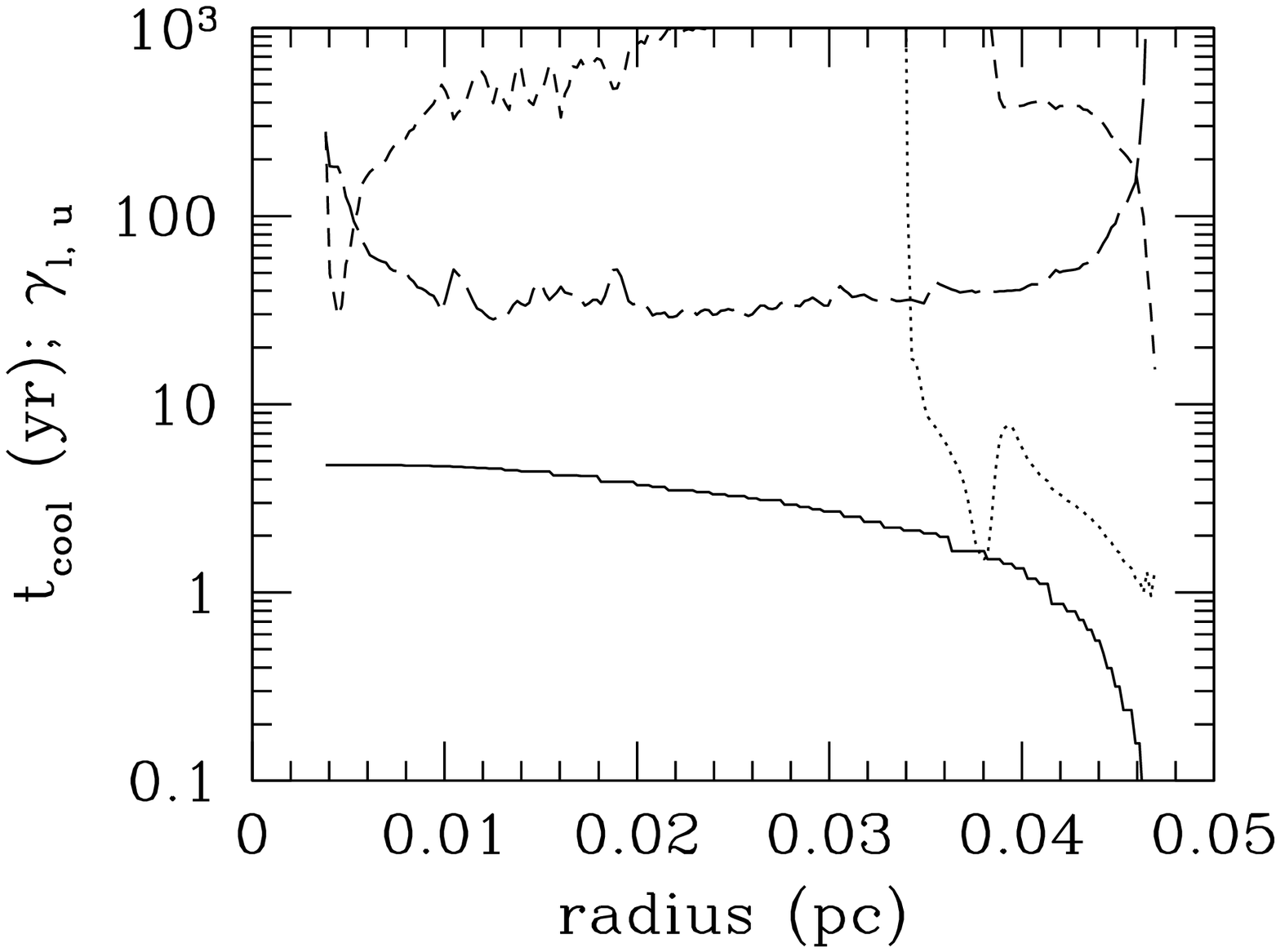}
\caption{The time since shock passage (solid line) and the non-thermal
  bremsstrahlung cooling time (dotted line) for electrons with
  relativistic $\gamma_u$ factor (long dashed line) corresponding to
  the synchrotron cooling time matching the time since shock
  passage. Also shown is the lower limiting $\gamma_l$ factor (short
  dashed line) below which electrons lose energy to plasmon excitation
  on a time scale shorter than shock passage. Emission is dominated by
  the layers immediately downstream of the shock front at
  0.047~pc. Shown for a $40\,{\rm M_\odot}$ hypernova 4.8~yr after the
  explosion, with $f_e=0.01$, $u_B=u_e$ and $p_e=2.0$ assumed.}
\label{fig:tc_gam}
\end{figure}

No definitive acceleration mechanism for non-thermal relativistic
electrons in supernova remnants has been identified, although
mechanisms involving the diffusive acceleration of particles by shock
waves, a form of first-order Fermi acceleration, are favoured. The
characteristic acceleration time, assuming Bohm diffusion, is
\begin{equation}
t_{\rm acc}\simeq\frac{r_g c}{u^2_{\rm sh}},
\label{eq:tacc}
\end{equation}
where $r_g=\gamma m_ec^2/eB\simeq1700(\gamma/B)$~cm is the gyroradius
of an electron with relativistic velocity factor $\gamma$ in a
magnetic field $B$, and $u_{\rm sh}$ is the shock velocity
\citep{2001RPPh...64..429M}. Requiring the acceleration time to be
shorter than the synchrotron cooling time
(equation~[\ref{eq:tau_sync}]) imposes the upper limit
\begin{equation}
\gamma_u^{\rm(sync)} < 1.2\times10^6 B^{-1/2}\left(\frac{u_{\rm
    sh}/c}{0.01}\right),
\label{eq:gam_u_sync}
\end{equation}
corresponding to a limiting electron energy of $\sim1$~TeV.

Downstream from the shock, at distances sufficiently far from the
shock front the electrons are not rapidly returned by turbulent eddies, the
electrons will cool in the absence of any other accelerating
mechanism. At a time $t^{\rm p-sh}$ after a shock, synchrotron cooling
will result in
\begin{equation}
\gamma_u^{\rm(p-sh,\, sync)} <
77.4B^{-2}\left[\frac{t^{\rm(p-sh)}}{\rm 10^7\,s}\right]^{-1}.
\label{eq:gam_u_p_sh}
\end{equation}

The electrons will also lose energy through scatters off other
electrons and the ions, generating non-thermal bremsstrahlung
radiation. The resulting cooling time is given by
\begin{equation}
(t_{\rm cool}^{\rm n-th\, brem})^{-1}=\frac{3}{\pi}\alpha\sigma_T
  c(n_{\rm H} + 3n_{\rm He})\left[\log(2\gamma) - \frac{1}{3}\right],
\label{eq:tcool_ntb}
\end{equation}
where $\alpha$ is the fine structure constant. Requiring
$t_{\rm cool}^{\rm n-th\, brem}>t_{\rm accel}$ imposes
\begin{equation}
\gamma_u^{\rm(n-th-b)}<4\times10^{10}\left(\frac{n_{\rm H}}{10^7\,{\rm
      cm^{-3}}}\right)^{-1}\left(\frac{u_{\rm sh}/c}{0.01}\right)^2 B,
\label{eq:gam_u_n_th-b}
\end{equation}
a less stringent constraint than given by synchrotron cooling for
$B>0.001\{(n_{\rm H}/10^7{\rm cm^{-3}})[0.01/(u_{\rm
    sh}/c)]\}^{2/3}$~G. For $\gamma$ factors of order 100, the
non-thermal bremsstrahlung cooling time approaches the synchrotron
cooling time.

Relativistic electrons will also lose energy through plasmon
excitation. The associated cooling time is
\begin{eqnarray}
(t_{\rm cool}^{\rm e-pl})^{-1}&=&\frac{3}{2}\sigma_T n_e
c\gamma^{-1}\left[\log\left(\gamma^{1/2}\frac{m_e
      c^2}{h \nu_p}\right)+0.216\right]\\\nonumber
&\simeq&2.99\times10^{-7}\frac{n_e}{10^7\,{\rm
    cm^{-3}}}\gamma^{-1}\\\nonumber
&&\times\left[(1/2)\log\left(1.89\times10^{25}\frac{10^7\,{\rm
    cm^{-3}}}{n_e}\gamma\right)+0.216\right]\,{\rm s^{-1}}.
\label{eq:tcool_epl}
\end{eqnarray}
In this case, the lowest energy electrons cool most quickly. Generally
$t_{\rm cool}^{\rm e-pl}>>t_{\rm acc}$ for the supernova remnants considered. The
requirement $t_{\rm cool}^{\rm e-pl}>t^{\rm p-sh}$ imposes the lower
limit on $\gamma$ of
\begin{equation}
\gamma_l^{\rm(p-sh,\, e-pl)}\simeq94.5\left(\frac{n_e}{10^7\,{\rm
      cm^{-3}}}\right)\left(\frac{t^{\rm p-sh}}{10^7\,{\rm s}}\right),
\label{eq:gam_l_e_pl}
\end{equation}
to within 5 percent accuracy for $t^{\rm p-sh}$ within an order of
magnitude of $10^7\,{\rm s}$.

\begin{figure}
\includegraphics[width=3.3in]{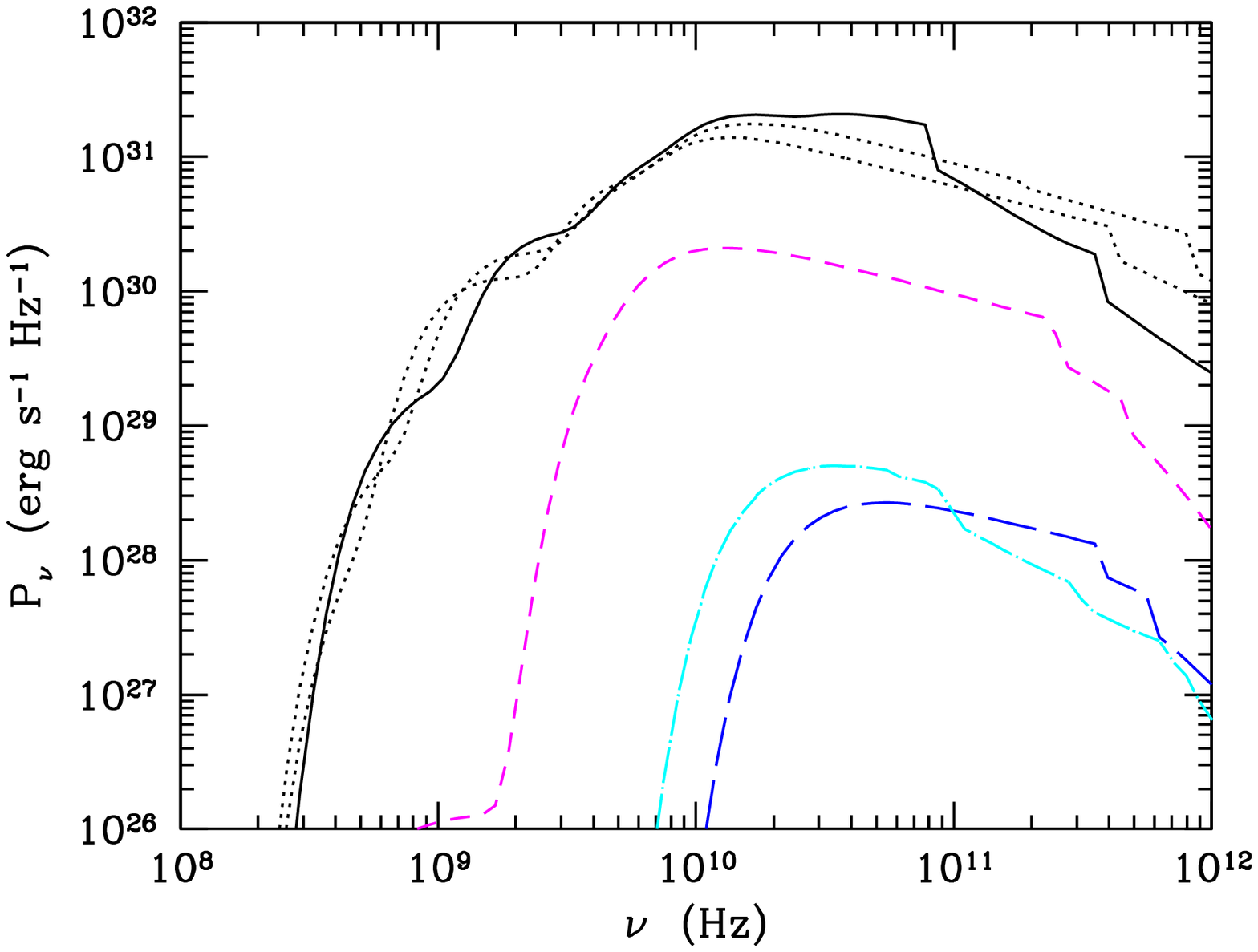}
\caption{The evolution of the synchrotron emission spectrum allowing
  for synchrotron self-absorption and free-free absorption, with
  $\gamma_l$ limited by loss to plasmon excitation and $\gamma_u$
  limited by synchrotron losses. Shown for a $40\,{\rm M_\odot}$
  hypernova, with $f_e=0.01$, $u_B=u_e$ and $p_e=2.0$ assumed, at
  times 4.7 (solid; black), 17.4 (short-dashed; magenta), 21.4 (long
  dashed; blue) and 32.5~yrs (dot-dashed; cyan) after the
  explosion. Also shown are models at 4.7~yrs after the hypernova,
  allowing $f_e$ to decrease due to cooling from an initial value of
  $f_e^i=0.05$ with $u_B/u_{\rm therm}=0.05$ held fixed (upper dotted
  curve; black) and with the initial value of $f_e^i=0.1$ and
  equipartition $u_B=u_e$ maintained (lower dotted curve; black).}
\label{fig:PSync_evol}
\end{figure}

The upper limit in the relativistic $\gamma$ factor for the electrons
is restricted primarily by synchrotron cooling, as illustrated in
Fig.~\ref{fig:tc_gam} for a $40\,{\rm M_\odot}$ hypernova in a
$1.2\times10^7\,{\rm M_\odot}$ halo at $z=17.3$, assuming $f_e=0.01$,
$u_B=u_e$ and $p_e=2.0$. Also shown are the upper and lower
relativistic factors $\gamma_u$ and $\gamma_l$ given by equating the
synchrotron cooling time and plasmon excitation energy loss time,
respectively, to the time since shock passage. By comparison, cooling
due to non-thermal bremsstrahlung losses is generally negligible. Most
of the interior of the supernova remnant would deplete its
relativistic electrons due to synchrotron and plasmon excitation
losses. Almost all the emission originates in a thin layer just
downstream of the shock front. The thickness $\Delta l_{\rm em}$ of
the layer may be estimated by requiring
$\gamma_l^{\rm(p-sh,\,e-pl)}<\gamma_u^{\rm(p-sh,\,sync)}$, giving
$\Delta t_{\rm em}/{\rm 10^7\,s} < 0.90(n_e/{\rm
  10^7\,cm^{-3}})^{-1/2}B^{-1}$, or
\begin{equation}
\Delta l_{\rm em}=u_{\rm sh}\Delta t_{\rm
  em}<2.7\times10^{15}\left(\frac{u_{\rm
    sh}/c}{0.01}\right)\left(\frac{n_e}{\rm
  10^7\,cm^{-3}}\right)^{-1/2}B^{-1}\,{\rm cm}.
\label{eq:dlem}
\end{equation}
Because of the steep rise in the thermal energy density just behind
the shock front, the emission is not much increased by relaxing
$\gamma_l$ to unity.

The evolution of the spectrum for a $40\,M_\odot$ hypernova remnant in
a $1.2\times10^7\,{\rm M_\odot}$ minihalo is shown at $z=17.3$ in
Fig.~\ref{fig:PSync_evol} for $f_e=0.01$ and $p_e=2$, with $\gamma_l$
and $\gamma_u$ restricted by energy losses. The curve at the time
$17.4$~yrs corresponds to density and temperature profiles shown in
\citet{2008ApJ...682...49W}. The emission power decays with time as
the shock weakens and the emitting region reduces in width according
to Eq.~(\ref{eq:dlem}). At 21.4~yrs after the hypernova, the emitting
zone has narrowed to $10^{15}\,{\rm cm}$. By 32.5~yrs, however, the
shock has weakened sufficiently for the emitting zone to broaden to
$3\times10^{16}\,{\rm cm}$, and the flux slightly recovers.

Because of energy losses by the relativistic electrons, a model with a
constant value for $f_e$ is an approximation. While the reduction in
$\gamma_u$ from synchrotron losses will little affect the total energy
content in relativistic electrons for $p_e>2$ because most of the
energy is at the lower energy end of the spectrum, it will reduce the
flux for $p_e<2$. For $p_e=2$ the energy is logarithmically
distributed. For an initial range $\gamma_l^i<\gamma<\gamma_u^i$, a
final range $\gamma_l<\gamma<\gamma_u$ will retain a fraction
$\log(\gamma_u/\gamma_l)/\log(\gamma_u^i/\gamma_l^i)$ of the original
energy. For $\gamma_l^i=1$, $\gamma_u^i=10^6$, $\gamma_l=50$,
$\gamma_u=300$, this gives a fraction 13\% of the original
energy. Thus an initial fraction $f_e^i=0.1$ would be reduced to
$f_e=0.013$. Since the magnetic energy density would not directly be
reduced by the losses, equipartition may not apply, in which case
$u_B>u_e$ may be possible, depending on the time to establish
equipartition. Spectra allowing for the energy loss since the time the
gas was shocked are also shown in Fig.~\ref{fig:PSync_evol} at 4.7~yrs
after the explosion for two alternative models, one assuming an
initial post-shock value $f_e^i=0.05$ and a post-shock magnetic energy
density to thermal energy density ratio fixed at $0.05$ and the second
assuming an initial post-shock value $f_e^i=0.1$ and a magnetic energy
density that maintains equipartition with the evolving energy density
of the relativistic electrons. Both these models well match the
constant $f_e=0.01$ model, so that the constant $f_e$ model is a good
approximation, noting that the value for $f_e$ is effectively an
average value.

At frequencies above $\sim100$~GHz, the models allowing for an
evolving $f_e$ due to cooling show an enhanced flux over the constant
effective $f_e$ model in Fig.~\ref{fig:PSync_evol}. In fact even in an
evolving $f_e$ model, the flux is expected to steepen at high
frequencies due to synchrotron losses. Under the assumptions of a
uniform shock velocity gas with a uniform magnetic field and diffusion
coefficient independent of electron momentum,
\citet{1987MNRAS.225..335H} solve the steady-state transport equation
for the population of relativistic electrons in the planar limit in
the presence of strong synchrotron cooling. They show a power-law
emission spectrum $\nu^{-0.5}$ will steepen to $\nu^{-1}$ above a
break frequency $\nu_b$ before cutting off above the maximum frequency
given by $\gamma_u$. The steepening is argued for on the general
grounds that the distance over which electrons cool out of the
distribution varies like $\gamma^{-1}$ for synchrotron cooling. The
break frequency depends on the specific model. They find $\nu_b\simeq
15\,B^{-3}[(u_{\rm sh}/c)/0.01]^2(\Delta l_{\rm em}/10^{15}\,{\rm
  cm})^{-2}$~GHz. For $B=0.1$~G and $\Delta l_{\rm em}=10^{16}$~cm,
this gives $\nu_b\simeq150$~GHz. While the diffusion coefficient is
generally expected to be proportional to the electron momentum
\citep{2001RPPh...64..429M}, the estimated break frequency is
comparable to that at which the constant effective $f_e$ model lies
below the evolving $f_e$ models and so may give a more realistic
representation of the spectrum. A more precise prediction would
require solving the relativistic electron transport equation, allowing
for a non-uniform shock speed and magnetic field as well as a
momentum-dependent diffusion coefficient; this is well beyond the
scope of this paper, and the predicted synchrotron spectrum would in
any case still depend on the assumed magnetic field properties and
initial electron spectrum. In the absence of a definitive theory of
relativistic electron acceleration and equipartition, the values
$f_e=0.01$ and $u_B=u_e$ are assumed for the estimates here, noting
that these are only meant to represent effective values.

As indicated by the above discussion, the results computed in this
paper will be quantitatively affected by several unknown factors,
including the relativistic electron acceleration mechanism, the
generation of turbulence at the shock front and its effects on the
relativistic electron energy spectrum and possibly on the propagation
of the shock front itself, and the subsequent transport of the
relativistic electrons as they cool due both to synchrotron and
plasmon excitation losses. By scaling the uncertainties based on
observed supernova remnants, we expect to minimize the effects of
these uncertainties on our predictions, but naturally cannot eliminate
them.

\end{appendix}

%%%%%%%%%%%%%%%%%%%%%%%%%%%%%%%%%
%\bigskip

\section*{acknowledgments}

We thank Philip Best, Jim Dunlop and Jeff Peterson for valuable
discussions. We thank the referee for useful comments. DJW was
supported by the Bruce and Astrid McWilliams Center for Cosmology at
Carnegie Mellon University. All ZEUS-MP simulations were performed on
Institutional Computing (IC) platforms at LANL (Coyote).

%
%%%%%%%%%%%%%%%%%%%%%%%%%%%%%%%%%

\bibliographystyle{mn2e-eprint}
\bibliography{apj-jour,mw}

\label{lastpage}

\end{document}